\documentclass{osa-article}

\journal{oe}

\articletype{Research Article}

\begin{document}

\title{Hot-spots and gain enhancement in a doubly pumped parametric down-conversion process}

\author{Ottavia Jedrkiewicz,\authormark{1,*} Erica Invernizzi,\authormark{2} Enrico Brambilla,\authormark{2} and Alessandra Gatti \authormark{3}}

\address{\authormark{1}Istituto di Fotonica e Nanotecnologie del CNR, Udr di Como, Via Valleggio 11, Como, Italy\\
\authormark{2}Dipartimento di Scienza e Alta Tecnologia, Università dell'Insubria, Via Valleggio 11, Como, Italy\\
\authormark{3}Istituto di Fotonica e Nanotecnologie del CNR, Piazza Leonardo Da Vinci 32, Milano, Italy}

\email{\authormark{*}ottavia.jedrkiewicz@ifn.cnr.it} 



\begin{abstract} 
We experimentally investigate the parametric down-conversion process in a nonlinear bulk crystal, driven by two non-collinear pump modes. The experiment shows the emergence of bright hot-spots in modes shared by the two pumps, in analogy with the phenomenology recently observed in 2D  nonlinear photonic crystals. By exploiting the spatial walk-off between the two extraordinary pump modes we have been able to recreate a peculiar resonance condition, reported by a local enhancement of the parametric gain, which  corresponds to a transition from a three-mode to a four-mode coupling. From a quantum point of view this opens the way to the generation of multimode entangled states of light such as tripartite or quadripartite states, in simple bulk nonlinear sources.
\end{abstract}

\section{Introduction}

Pulsed high-gain parametric down-conversion (PDC) in standard nonlinear crystals is widely used for the generation of bright squeezed vacuum, featured by perfect photon-number correlations between twin-beams  \cite{Jedrkiewicz2004,Bondani2007,Brida2009,Iskhakov2009,Agafonov2010,Perez2015}.  This non-classical state of light has interesting applications in different fields such as quantum imaging \cite{Jedrkiewicz2004,Brida2010}, sensing\cite{Lopaeva2013}, or macroscopic entanglement\cite{Iskhakov2012}. In high-gain PDC, down-converted radiation is exponentially amplified along the medium, and appears as a bright conical emission in well-defined spatial and temporal modes ruled by the phase-matching conditions. It has been shown that although walk-off effects, originating from the difference between the group and phase velocities, limit the efficiency of nonlinear optical interactions, they can be turned into useful tools for shaping the emission \cite{Perez2015}. 

On the other hand, nonlinear photonic crystals (NPC), whose nonlinear response is artificially modulated according to a 2D pattern \cite{Arie2009}, offer a high degree of flexibility for engineering the properties of optical parametric processes because of the
multiplicity of vectors of the nonlinear lattice providing
quasi phase-matching. These photonic crystals have shown interesting potentialities as monolithic sources of path-entangled photonic states \cite{Gong2012,Jin2013,Megidish2016}, but even if they may provide novel compact schemes for continuous-variable quantum technologies \cite{Gong2016,Jedrkiewicz2018,Gatti2018,Gatti2020a}, they are often difficult to realize, necessitating of lengthy poling procedures.
 Of particular interest for quantum technologies is the possibility, recently outlined \cite{Jedrkiewicz2018,Gatti2018,Gatti2020a}, of using NPC as sources of three-mode or four-mode parametric coupling, in contrast to the more common  two-mode coupling typical of the PDC  process in standard configurations. 
 
The aim of this work is to experimentally reproduce this multimode parametric coupling in a traditional (non engineered) nonlinear source, by using a spatially multimode beam to pump the process. The idea, explored in detail by some of us in a strictly related theoretical work \cite{Gatti2020}, comes from the observation that the same effective nonlinear polarization term  can be obtained either by having a transverse modulation of the nonlinearity of the medium and using a spatially homogeneous pump,  as done in NPCs, or by employing a uniform nonlinear medium illuminated by a spatially modulated pump.

The experiment has been implemented by considering the type I process in a Beta-Barium-Borate (BBO) crystal. A transverse spatial modulation of the pump, in the form of a stripe pattern, is generated by interfering two well distinguishable laser beams in a Mach-Zehnder type scheme, with the possibility of controlling the tilt angle between the two modes. Each pump mode (denoted by p$_{1}$ and p$_{2}$ respectively) generates in the far field a family of cones of parametric radiation collinear with its direction. The intersection zones define an ensemble of modes shared by the two pumps, indistinguishably populated by the photons generated by pump p$_{1}$ or by pump p$_{2}$.  In analogy to what observed in NPC \cite{Jedrkiewicz2018}, we expect that in the high-gain regime these two possibilities sum up coherently, leading to a local increase of the parametric gain. The same local increase of the parametric gain affects the pair of modes coupled to each shared mode. As predicted in \cite{Kolobov2010} and shown in more detail in \cite{Gatti2020}, the doubly-pumped BBO scheme realizes in these conditions a three-mode entangled state (more precisely, a multiplicity of independent 3-mode states corresponding to each triplet of modes).
In this work we indeed experimentally demonstrate, and quantitatively analyze, such a local gain enhancement, which results in an unusual far-field distribution of the down-converted light, in the form of three bright stripes of hot-spots against a more diffuse background.

In addition to the 3-mode coupling, this works demonstrates the possibility to make a transition to a 4-mode coupling, characterized by a Golden Ratio enhancement of the parametric gain in the hot-spots. Such a transition was observed in a NPC source \cite{Jedrkiewicz2018,Gatti2018} as a result of a spatial resonance taking place between the transverse modulation of the nonlinear lattice and the phase-profile of a single-mode pump beam. In our case, somehow surprisingly, a completely analogue transition can be realized by exploiting the spatial walk-off between the two pump modes, and an interpretation of this condition can be given in terms of a superposition of the career Poynting vector, identifying the direction of propagation of the energy flux, with either one pump mode or the other  \cite{Gatti2020}.
In the experiment, such a \textit{resonance} condition (as we shall call it from now on) is reached by continuously rotating the crystal in the plane parallel to its input facet, i.e. the plane orthogonal to the crystal principal plane containing the optical axis.
For a given tilt angle between the two pump modes, we indeed show that there exist particular rotation angles of the crystal such that two of the three far-field stripes of hot-spots coalesce, with a corresponding increase of their intensity. The latter, originating from a local enhancement of the parametric gain by a factor equal to the Golden Ratio, is in accordance with what predicted by the theoretical description of the 4-mode coupling \cite{Gatti2018,Gatti2020}.

The paper is organized as follows: in section 2 we present the experimental set-up used for this work; in section 3 we introduce some concepts and results giving a theoretical insight of the mode coupling involved in the parametric down-conversion process pumped by a spatially modulated beam pattern resulting from the interference of two relatively tilted pump beams. In particular we present the expected spatial distribution of the coupled and shared modes of the PDC radiation in different conditions of crystal orientation. In Section 4 we describe the diagnostics used to detect and to spatially and spectrally characterize the radiation hot spots, and we present the experimental results. We also discuss the observed gain enhancement at \textit{resonance}. The conclusions are presented in section 5.

\section{The experimental set-up}

Figure 1 shows a scheme of our experimental set-up. The pulsed UV pump at 352 nm is obtained as the third harmonic of an amplified Q-switched Nd:Glass laser, delivering 10 Hz s-polarized pulses of 1.5 ps duration at 1055 nm wavelength (see supplemental material for details).
The two interfering UV pump beams, slightly non-collinear, are obtained by using a Mach-Zehnder type interferometer, as shown in Fig. 1a. An external tilt angle $\vartheta_{p}$ between the two beams is generated in the transverse direction (in the horizontal laboratory plane) by slightly rotating one of the two mirrors M. The same mirror is mounted on a micrometric translation stage used to correct the delay due to the optical path difference of the UV pulses in the two arms of the interferometer. The output 50/50 beam-splitter (BS$_{2}$) is also mounted on a micrometer translation stage used to compensate the M mirror displacement, so to guarantee the superposition of the two beams at the very output of the interferometer a few centimeters after BS$_{2}$, in the focal plane of lens L$_{2}$ (f$_{2}$=25 cm). Note that this lens is part of a telescopic system used to demagnify the input UV beam (as illustrated in Fig. 11 of the supplemental material). The tilt angle between the two pump modes can be in principle derived from their spatial interference pattern observed by a CCD camera in that plane. Fig. 1b shows an example of such a fringe pattern, from which a tilt angle $\vartheta_{p} \simeq 0.7^{\circ}$ could be estimated. In order to resolve the hot-spot distribution of the emitted PDC radiation, slightly larger angles are needed. To this end, the fringe pattern is demagnified through an imaging system by a factor $p/q=1.5$ with one lens (L$_{3}$) having focal length f$_{3}$= 15 cm, and the interferometer alignment has been progressively optimized during the data acquisition, till reaching $\vartheta_{p} \simeq 2^{\circ}$. Notice that in the imaging plane of L$_{3}$ the fringes cannot be resolved due to an insufficient spatial resolution of the detector, while the transverse profile of the spatially modulated pump has a full width at half maximum (FWHM) of about 350 $\mu m$.

\begin{figure}[ht]
\centering\includegraphics[width=10cm]{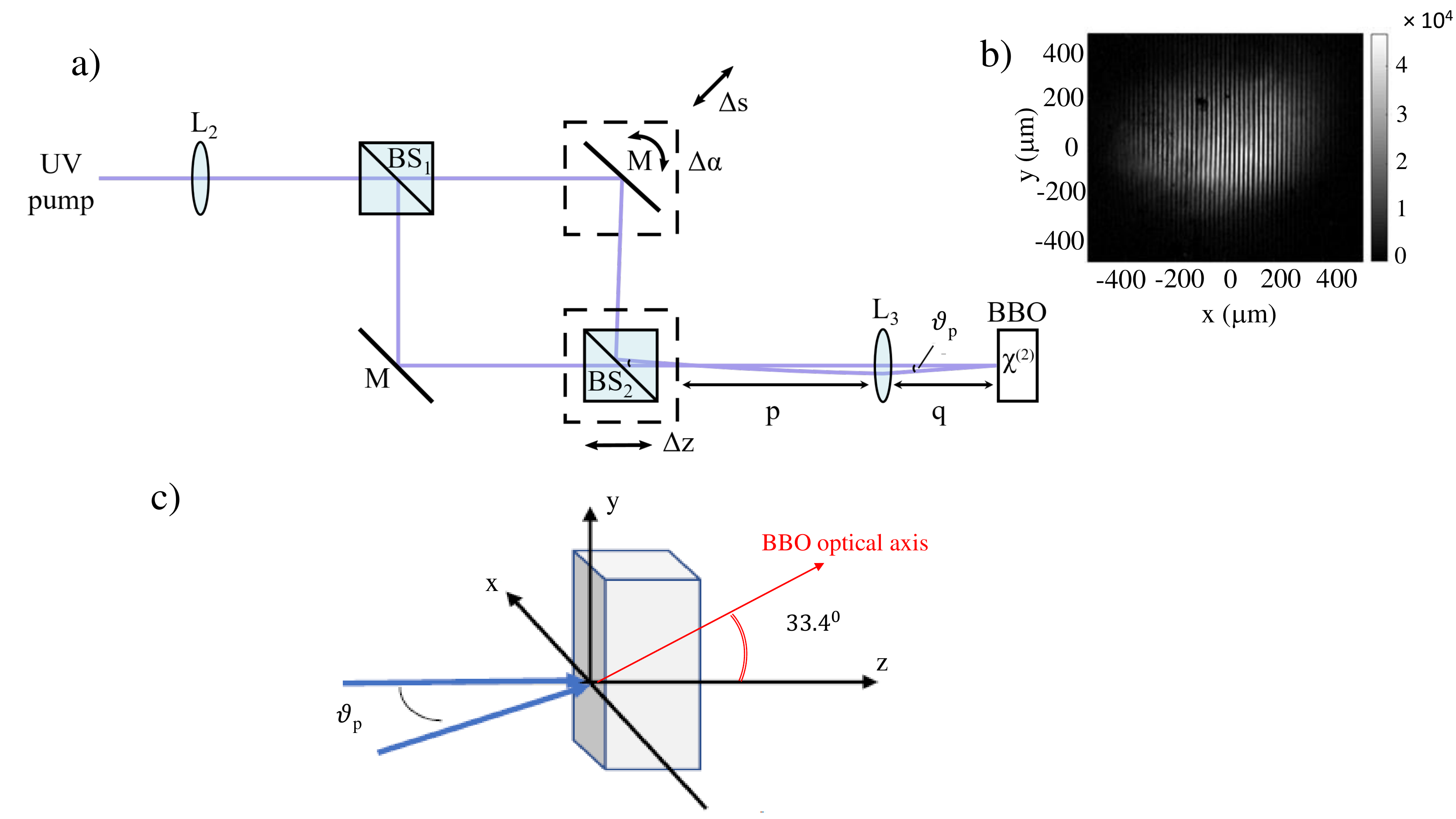}
\caption{Experimental set-up. (a) Scheme of the Mach-Zehnder type interferometer splitting the UV pump into two beams tilted in the horizontal plane (xz). The figure is not in scale, the focal lengths of L$_{2}$ and L$_{3}$ are respectively 25 cm and 15 cm; p= 37.5 cm and q=25 cm. (b) Stripe pattern generated by the interference of the two pump modes, recorded by a CCD camera after BS$_{2}$ in the focal plane of L$_{2}$. (c) Geometry of the BBO crystal, showing the orientation of the optical axis.}.
\end{figure}

The entrance facet of the BBO crystal is initially placed at 25 cm from L$_{3}$, in its imaging plane, where the two UV pumps are superimposed. The crystal is 4 mm-long and cut at 33.48$^{\circ}$ for type I collinear PM when pumped at $\lambda_{p}=352nm$. The crystal is mounted on a stage allowing its orientation in all directions, and in particular a precise rotation in the (xy) plane. Its position along z is then finely adjusted by optimizing the intensity of the PDC signal in the double pumping configuration.

\section{Theoretical prediction}

A theoretical study of the scheme, including an analysis of the multipartite entangled state generated,  is presented in the related paper \cite{Gatti2020}. In the following,  we limit the discussion to results useful for understanding the experimental findings that will be presented in Section 4. 
Numerical simulations of the PDC process have also been performed. These (see methods of \cite{Jedrkiewicz2018}) are fully 3D +1 simulations, based on the nonlinear coupled propagation equations of the signal and pump field envelopes presented in \cite{Gatti2020}. They have been carried out by considering two input  Gaussian pump pulses, tilted one with respect to the other by $2^\circ$ in the transverse plane, each one having a waist of 297 $\mu m$  (corresponding to 350 $\mu m$ FWHM in the spatial intensity profile), and a pulse duration of 1.2 ps. The PDC frequency bandwidth considered in the simulations is 600-850 nm.  

In order to understand the analogy between our experiment and the process of parametric generation in 2D nonlinear photonic crystals, we recall that in such materials the nonlinear response is artificially modulated typically via ferroelectric poling, according to a 2D pattern. We can observe that for a 2D photonic crystal the second order nonlinear polarization term acting as a source in the propagation wave equation for the signal field can be written in the direct space as

\begin{equation}
P_{s}^{NL}(x,y,z,t)\propto \varepsilon_{0} \chi^{2}(x,z)E_{p}(x,y,z,t)E_{i}^{*}(x,y,z,t).
\label{PNLsfot}
\end{equation}

\noindent
In the undepleted pump case the pump profile remains constant during propagation. The same nonlinear polarization term can be obtained by a transverse spatial modulation of the second-order non linear response of the medium $\chi^{2}$, as in the 2D photonic crystal, or keeping the susceptibility constant and spatially modulating the transverse profile of the  pump field $E_{p}(x,y,z,t)$. For this reason, in this work, we use two pump beams slightly tilted one with respect to the other in the horizontal $(x,z)$ plane of the laboratory, forming by interference an intensity transverse pattern featured by stripes (i.e. a spatial modulation along $x$). Each pump generates a set of parametric fluorescence cones, the two sets being not coaxial.

In order to describe the experiment, we adopt a reference frame $(x,y,z)$ corresponding to the laboratory reference frame, in which the two pump beams 
$p_1$ and $p_2$ propagate in the horizontal $(x,z)$ plane, with a small tilt angle $\theta_{p}$ along the transverse $x$ axis (notice that in the following   $\theta_{m} $  denote all the internal angles, $m$ referring to the pump or signal modes, while $\vartheta_{m} $ are the corresponding external angles).
In the analytical calculation, we  assume that the pumps are plane-wave modes, 
with wave-vectors 
\begin{eqnarray}
&\boldsymbol{k_{p1}} &=Q_{1x}\hat{\boldsymbol{e}}_{x}+k_{1z}\hat{\boldsymbol{e}}_{z} \\
&\boldsymbol{k_{p2}}&=Q_{2x}\hat{\boldsymbol{e}}_{x}+k_{2z}\hat{\boldsymbol{e}}_{z},
\label{vettori d'onda 2 pompe}
\end{eqnarray}
so that  $\boldsymbol{Q}_{j}=Q_{jx}\hat{\boldsymbol{e}}_{x}$ (j=1,2)  are their transverse wave vectors.

 In the experiment we  choose the wave vector $\boldsymbol{k_{p1}}$ of pump p$_{1}$ to be collinear to the direction $z$, i.e. $Q_{1} =0$.  The crystal is initially placed in such a way that the reference frame $(x',y',z)$ parallel to the crystal facets coincides with the laboratory frame $(x,y,z)$. For practical reasons, we  investigated the existence of the "resonance" condition by keeping fixed the propagation direction of the pump beams and by rotating the crystal in the $(x,y)$ plane (as shown for instance in Figure 4). We call $\beta$ the angle formed between the $(x',y')$ axes of the crystal reference frame and $(x,y)$. Such a rotation of the crystal implies an azimuthal rotation of the optical axis of the crystal (which lies in the plane $(y',z)$, see Fig.1)  around the  $z$ axis. This in turn induces a strong change of the refractive index  of the tilted pump p$_{2}$ when varying $\beta$. The larger the rotation  $\beta$,  the larger the variation of the wave-number  $k_{p2} = |\boldsymbol{k_{p2}}|$ of pump 2, depending on its propagation direction $\theta_{p}$ with respect to $z$.

 In the type I (e-oo)  degenerate  phase-matching (PM) configuration of the PDC process, the signal and idler fields have the same polarization and are generated in a  large bandwidth around the degenerate frequency $\omega_{s}=\omega_{p}/2$ ($\omega_{p}$ being the pump frequency). Assuming a monochromatic pump, because of energy conservation the signal and idler photons  generated in each elementary down-conversion process are emitted  at conjugate frequencies  $\omega_s +\Omega_s$ and $\omega_s -\Omega_S$, where $\Omega_s$ denotes a frequency shift from the career. 
 Defining as $\boldsymbol{q}_{j}=q_{jx}\hat{\boldsymbol{e}}_{x}+q_{jy}\hat{\boldsymbol{e}}_{y}$ ($j$=s,i) the transverse components of their wave vectors,  the transverse phase-matching conditions for the two processes are given by 
\begin{eqnarray}
&\boldsymbol{q}_{s}+\boldsymbol{q}_{i}=\boldsymbol{Q}_{1} \label{PM trasversop1}\\
&\boldsymbol{q}_{s}+\boldsymbol{q}_{i}=\boldsymbol{Q}_{2}, \label{PM trasversop2}
\end{eqnarray}
\noindent
while the longitudinal PM is determined by the momentum conservation along z, and can be written as:
\begin{eqnarray}
&D_{1}(\boldsymbol{q}_{s},\Omega_{s})=k_{sz}(\boldsymbol{q}_{s},\Omega_{s})+k_{sz}(-\boldsymbol{q}_{s}+\boldsymbol{Q}_{1},-\Omega_{s})-k_{1z}\simeq 0  
\label{PM long D1}\\
&D_{2}(\boldsymbol{q}_{s},\Omega_{s})=k_{sz}(\boldsymbol{q}_{s},\Omega_{s})+k_{sz}(-\boldsymbol{q}_{s}+\boldsymbol{Q}_{2},-\Omega_{s})-k_{2z} \simeq 0 \label{PM long D2}
\end{eqnarray}
where energy and  transverse momentum conservation have been incorporated.

\begin{figure}[htbp]
\centering\includegraphics[width=10cm]{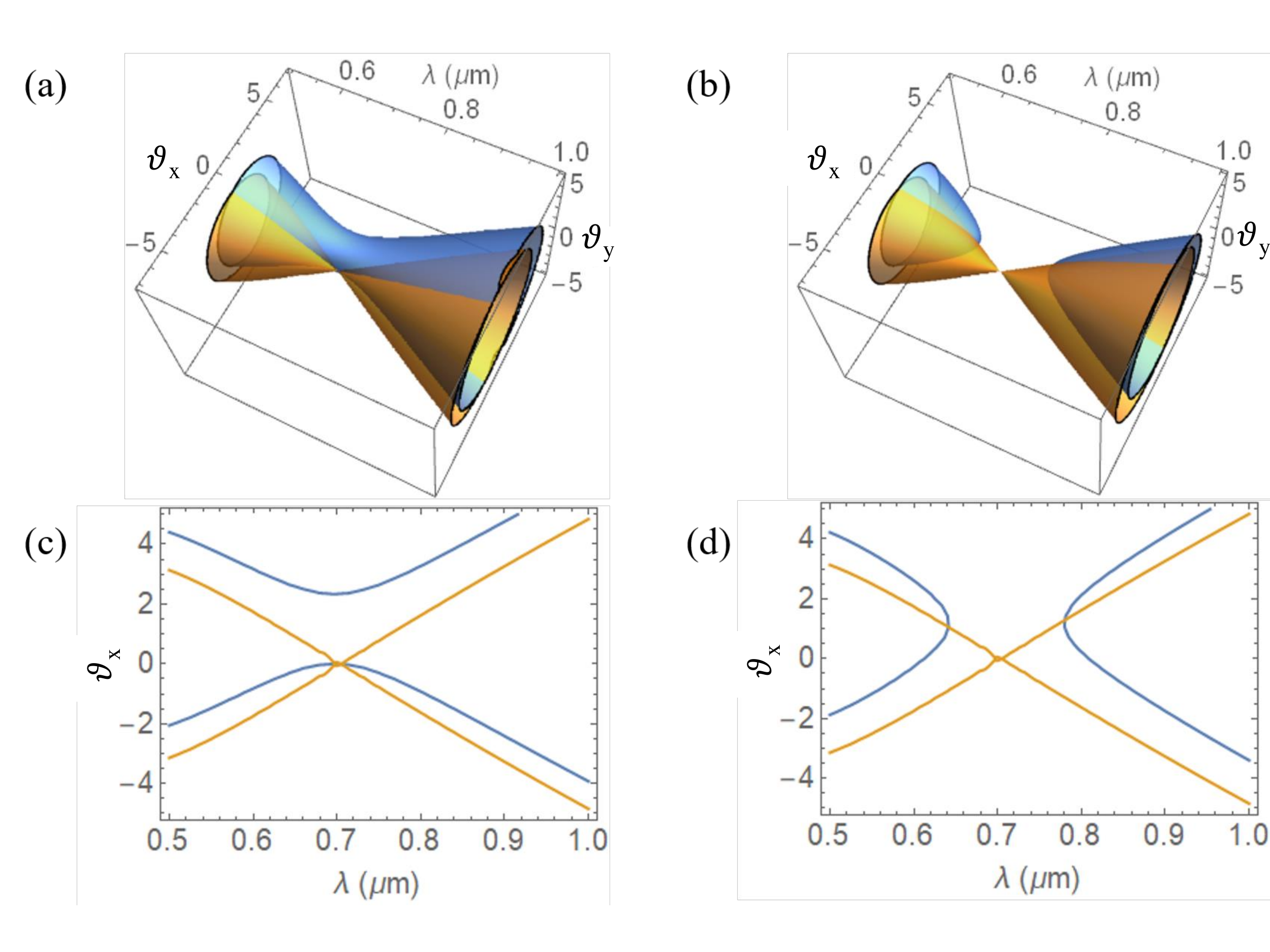}
\caption{Phase-matching surfaces of the PDC process with two pumps at \textit{resonance} condition (see next Section) corresponding to $\beta=7^{\circ}$ (a) and $\beta=-8.8^{\circ}$ (b). The phase-matching curves (c) and (d) are the sections of the surfaces (a) and (b), respectively,  in the plane of the pumps ($\vartheta_{y}=0^{\circ}$). The pump p$_{1}$ (orange) propagates along the $z$ axis of the crystal, while p$_{2}$ (blue) is tilted along  $x$  by an external angle of $2^{\circ}$. In the case of (a,c) the PM for p$_{2}$ is a non collinear type PM around degeneracy, while in the case of (b,d) the PM is non degenerate.}
\end{figure}

Equations \eqref{PM long D1} and \eqref{PM long D2} define two surfaces in the Fourier space of the down-converted light, that we call $\Sigma_{1}$ and $\Sigma_{2}$ (see Fig.2). They describe  the phase-matched modes for the PDC process associated with each pump, and correspond  
in the angular and spectral domain, to a standard conical emission around each pump. The interesting point is that these surfaces are in fact intersecting in some regions and the two processes coexist inside the nonlinear crystal. 

\subsection{Shared-coupled modes and transition to resonance} 

The modes lying at the geometrical intersection $\Sigma_{1} \cap \Sigma_{2}$ are particular because here the PM is simultaneously realized for both pumps.  Therefore a photon appearing in one of these \textit{shared} modes (as we shall call it) originates from either pump 1 or pump 2, indistinguishably. As a consequence,  its twin idler is emitted  in either one of the two coupled modes. For the signal field the \textit{shared} mode geometrical distribution in the angular and spectral domain is defined by

\begin{equation}
D_{1}(\boldsymbol{q}_{s},\Omega_{s})=D_{2}(\boldsymbol{q}_{s},\Omega_{s}).
\label{definizioneshared}
\end{equation}
Since each \textit{shared} mode is coupled to two modes at the conjugate frequency, the doubly pumped PDC process is generally featured by  triplets of modes that evolve according to  3 coupled propagation equations,   and whose state is a tripartite entangled state, as described in \cite{Gatti2020}. Such 3-mode coupling is schematically depicted in Fig.3a. 

\begin{figure}[ht]
\centering\includegraphics[width=12cm]{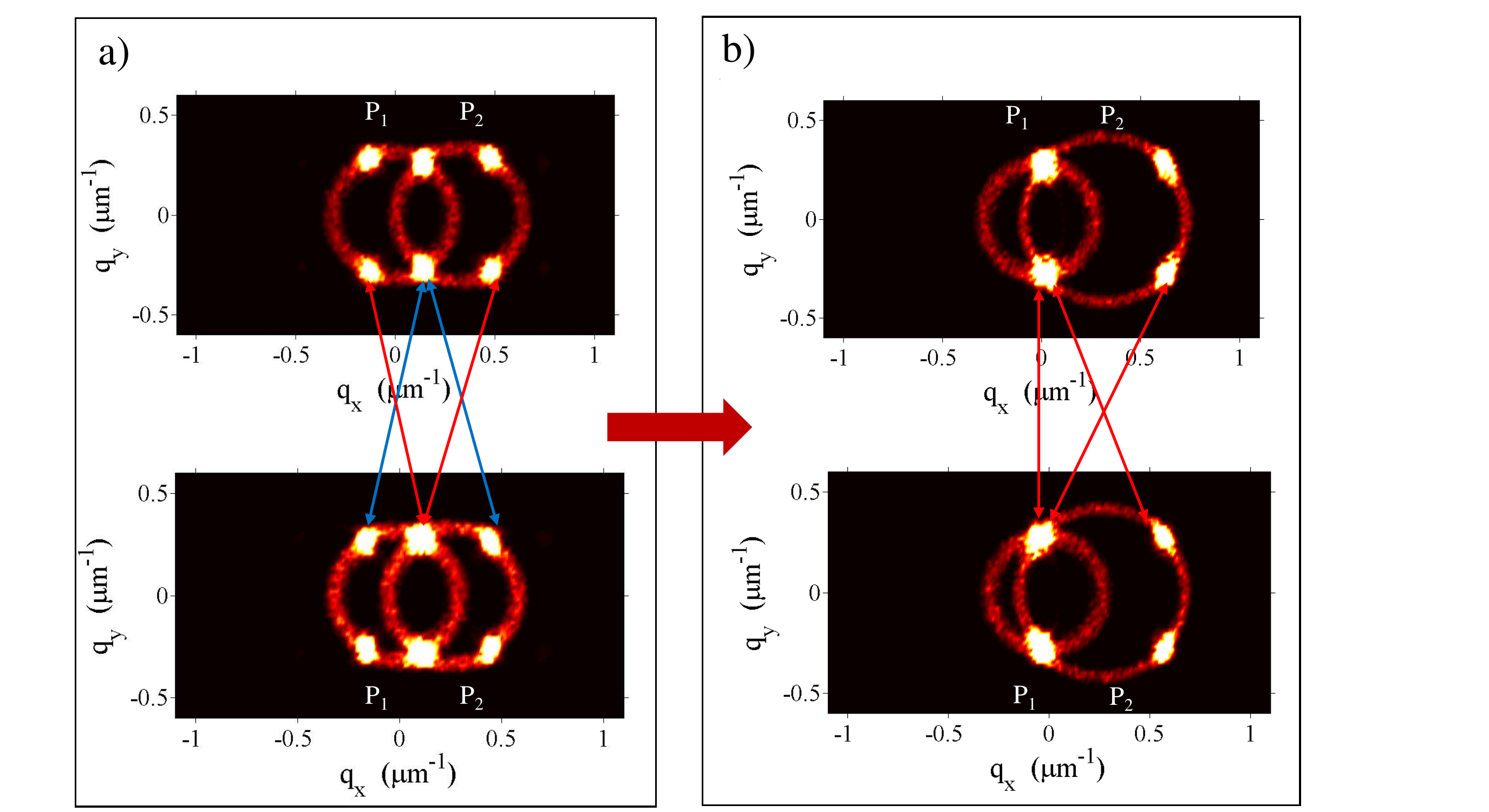}
\caption{Numerical simulations of the PDC light distributions in the  (q$_{x}$,q$_{y}$) plane, illustrating the three-mode (a) and the four-mode coupling (b) for two  conjugate wavelengths.  In (a) $\beta=0^{\circ}$, and the plot   highlights the coupling occurring inside two independent triplets of modes,  respectively defined by the \textit{shared} signal mode and the related \textit{coupled} idler modes (indicated by the red arrows), and by the \textit{shared} idler modes and the related \textit{coupled} signal (blue arrows). In (b)  $\beta=7^{\circ}$: at resonance the two triplets of modes merge into four modes, and the \textit{shared} modes superimpose to the coupled mode generated by the pump p$_{1}$. The simulation considered two small conjugated bandwidths, given by $\Delta\lambda_{s}=$635-645 nm and $\Delta\lambda_{i}$=775-789 nm (a), and $\Delta\lambda_{s}=$640-650 nm and $\Delta\lambda_{i}$=768-782 nm (b). The gain is $gl_{c}$=6 with a crystal length $l_{c}$=4mm. The intensity is reported in arbitrary units.}
\end{figure}

Mathematically, the angular-spectral position of  \textit{shared} and \textit{coupled} modes  are determined by the condition $D_{1}(\boldsymbol{q}_{s},\Omega_{s})=D_{2}(\boldsymbol{q}_{s},\Omega_{s})\simeq0$. Using the results in \cite{Gatti2020}, the angular positions of these modes in the $x$ direction are given by 

\begin{align}
\theta_{0x}(\Omega_{s}) &=\dfrac{\theta_{p1}+\theta_{p2}}{2} + \dfrac{\Delta k_{p}}{\Delta Q_{p}} \dfrac{k_{s}(-\Omega_{s})}{k_{s}(\Omega_{s})}
\\
\theta_{1,2x}(\Omega_{s}) &=\dfrac{\theta_{p1}+\theta_{p2}}{2} \pm\dfrac{\theta_{p1}-\theta_{p2}}{2}\dfrac{k_{p}}{k_{s}(\Omega_{s})}-\dfrac{\Delta k_{p}}{\Delta Q_{p}} \label{sistemaSeCeq2}
\end{align}

\noindent
where the subscript $0$ refers to the \textit{shared} mode and the subscripts $1,2$ refer to the two \textit{coupled} modes; while in the $y$-direction $\theta_{0y}(\Omega)=\theta_{1,2y}(\Omega)$. 
The parameter
\begin{eqnarray}
\dfrac{\Delta k_{p}}{\Delta Q_{p}}=\dfrac{k_{p2}-k_{p1}}{Q_{2}-Q_{1}}
\label{rate}
\end{eqnarray}
measures the rate of variation of  the  pump wave-numbers with their transverse tilt. This parameter appears inside the two equations with opposite signs, so that by  continuously varying it, one of the two side modes may  superimpose to the central \textit{shared} mode of the same frequency, to which it was originally uncoupled. In our experiment the parameter is indeed varied by rotating the crystal in the transverse plane. Alternatively, one could keep the crystal fixed and change the tilt angle between the two pumps until 
\begin{equation}
  \dfrac{\Delta k_{p}}{\Delta Q_{p}} =\mp \frac{\theta_{p2}-\theta_{p1}}{2}  
\end{equation}
which represents the {\em resonance condition}: as demonstrated in \cite{Gatti2020},   to a very good approximation it does not depend on the frequency, and it is achieved simultaneously in a huge bandwidth around the degenerate wavelength.

This dynamics is illustrated in Figures 3 and 4. 
 Figure 3 is obtained from the numerical simulations of the process, and plotted for a restricted spectral bandwidth around two conjugate wavelengths so to highlight the hot-spots. The plots show the transition to resonance in the case where the \textit{shared} modes superimpose to the \textit{coupled} modes generated by the pump p$_{1}$.  In these conditions two independent   triplets of modes merge into 4 modes, whose coupling is schematically shown by the red arrows in Fig.3b, giving rise to an interesting quadripartite entangled state \cite{Gatti2020}.

\begin{figure}[ht]
\centering\includegraphics[width=12cm]{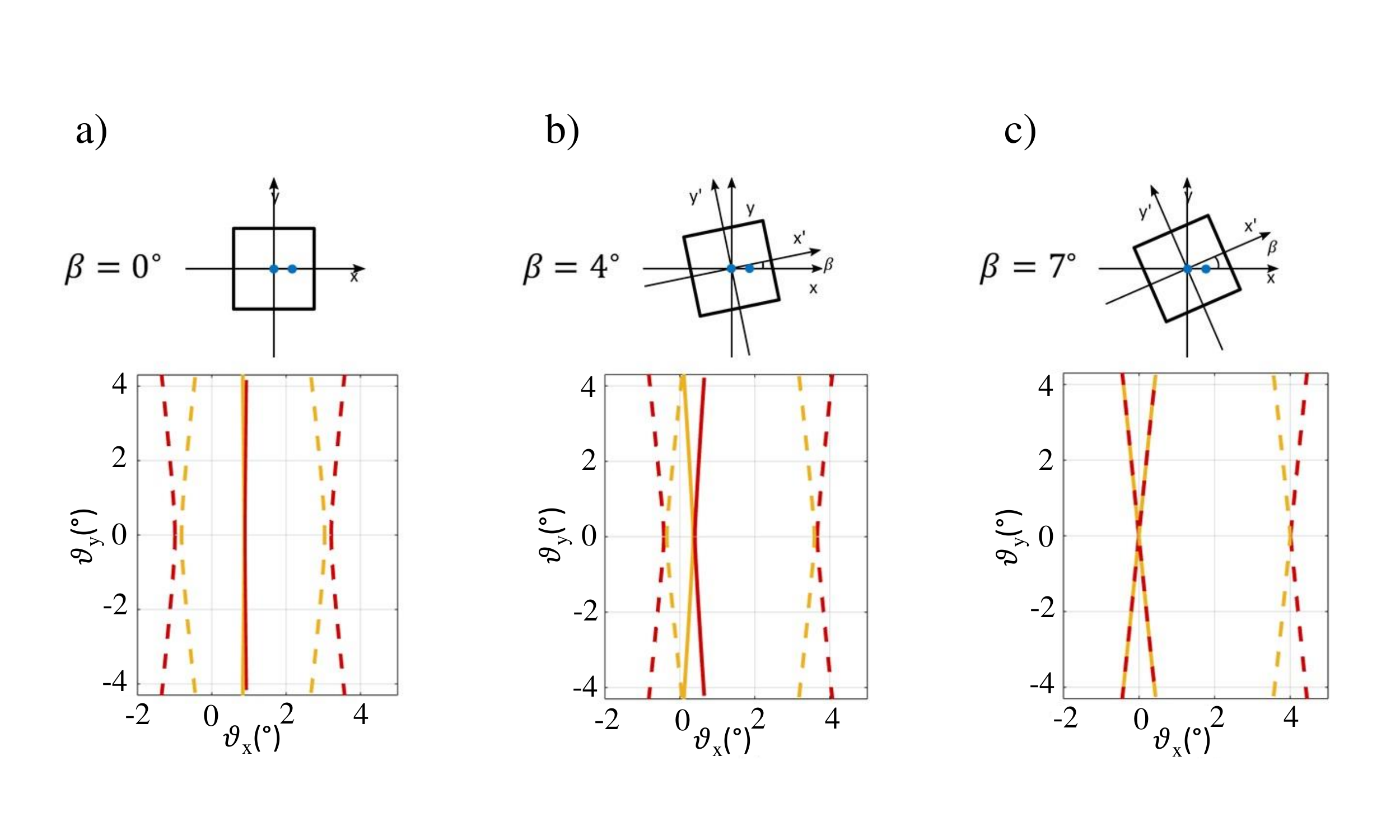}
\caption{Angular positions of the shared and coupled modes, for increasing values of  $\beta$, showing the transition to resonance. (a)  $\beta=0^\circ$, (b) $\beta=4^\circ$, (c) $\beta=7^\circ$ (resonance).
Solid and dashed lines are associated respectively with all shared and coupled modes displayed over an angular range of $\pm 4^\circ$ corresponding to a spectral bandwidth of 550-850 nm. Different colours are associated with independent branches of triplets, so that solid and dashed lines of the same colour plot the position of shared modes and their coupled ones, respectively. While changing $\beta$ the modes are progressively displaced and at resonance (c) shared and coupled modes superimpose. The upper row shows the corresponding crystal configuration. The two pumps lie in the (x,z) plane and are tilted by $\vartheta_{p}=2^\circ$ (external angle).}
\end{figure}
 
 \noindent
Figure 4 instead plots  the angular distribution of the  \textit{shared} and \textit{coupled}  modes without resolving the wavelengths, corresponding to the far-field distributions observed in the experiment. 
In this case, the hot-spots associated with shared and coupled modes merge into continuous bands. The curves are obtained  from Eqs. (9) and (10), by using the Sellmeier formulas in \cite{Kato86} and are  plotted for values $\beta$ growing till the resonance condition $\beta = 7^\circ$ is reached. We note that at resonance not only the \textit{shared} modes superimpose to one branch of the \textit{coupled} modes, but they are approximately collinear with a pump mode. \\
The 3D shape of the phase-matching surfaces at resonance, as well as their spectral angular distributions in the ($\lambda, \vartheta_x$) plane, can be appreciated in Fig. 2,  for $\beta=7^{\circ}$ in (a) and (c), and $\beta=-8.8^{\circ}$ in (b) and (d)),  corresponding to the two resonances.  
\par 
The parametric gain enhancement associated with the hot-spots can be evaluated from the following arguments:\\
- Outside resonance,  the 3-mode coupling has been demonstrated \cite{Gatti2020} equivalent to a single parametric process of gain $g \propto \sqrt{2 I_p}$, where $I_{p}$ is the intensity of each pump mode, assuming that their energy is perfectly balanced.  In the high-gain, where the intensity grows as $\sim \sinh^2 (g l_c) $ we thus  expect a local enhancement of the intensity at the hot-spot positions $\sim I_p^{\sqrt 2}$, with respect to the 2-mode fluorescence from each pump. \\
- At resonance, the 4-mode coupling is instead equivalent to two parametric processes of gains respectively given by $\phi$ and $1 \over \phi $, where $\phi=1.618... $ is the Golden Ratio. Since in the high-gain regime the process with larger gain prevails over the other one, we expect a local intensity enhancement $I_p^{\phi}$. The results for two balanced pumps are indeed completely equivalent to those predicted and demonstrated for the NPC \cite{Gatti2018,Jedrkiewicz2018}.


\section{Experimental results}

The transition to resonance described in the previous section and the associated local intensity enhancement of the radiation, are experimentally   detected and studied by spatially and spectrally resolving the \textit{shared} and \textit{coupled} modes and the PM curves associated with the two PDC processes.
The set-up used for the diagnostics is described  below.

 A dichroic mirror (M$_{4}$) placed  just after the crystal (see Fig. 5) reflects away the pump and a coloured filter with a 90-95$\%$ transmission in the bandwidth 450nm-2000nm is used to absorb UV pump residues. Neutral density filters (ND) are used during the measurements to attenuate the radiation intensity before detection. 
The spatial intensity distribution of the PDC radiation emitted by the BBO crystal is analyzed in the far-field, with a high efficiency, 16 bit, CCD camera (Andor) cooled down to -50$^{\circ}$C, placed in the focal plane of a lens (see Fig. 5a). The CCD sensor constituted by 1024 x 255 pixels of size 26 $\mu m^{2}$ each, is featured by a maximum quantum efficiency of 95$\%$ in the visible range decreasing towards the infrared region of the spectrum (with a quantum efficiency of about 85$\%$ at degeneracy for $\lambda_{deg}=704 nm$).
Spectrally resolved intensity distributions are recorded by means of the same CCD coupled to an imaging spectrometer (Lot Oriel) with a 200 $\mu m$ thin vertical slit selecting a portion of the PDC intensity pattern in the focal plane of the far-field lens L$_{4}$ with f$_{4}$= 5 cm (see Fig. 5b). Note that in order to spectrally analyze the radiation emitted in the horizontal plane (i.e. the plane containing the two pumps) the intensity pattern is rotated by 90$^{\circ}$ in the (xy) plane by means of a couple of mirrors (not shown in the figure) placed between the far-field lens and the slit. A mechanical shutter (S) controlled by the CCD camera software is used to select single pulses or to acquire the images with a suitable integration time.

\begin{figure}[ht]
\centering\includegraphics[width=10cm]{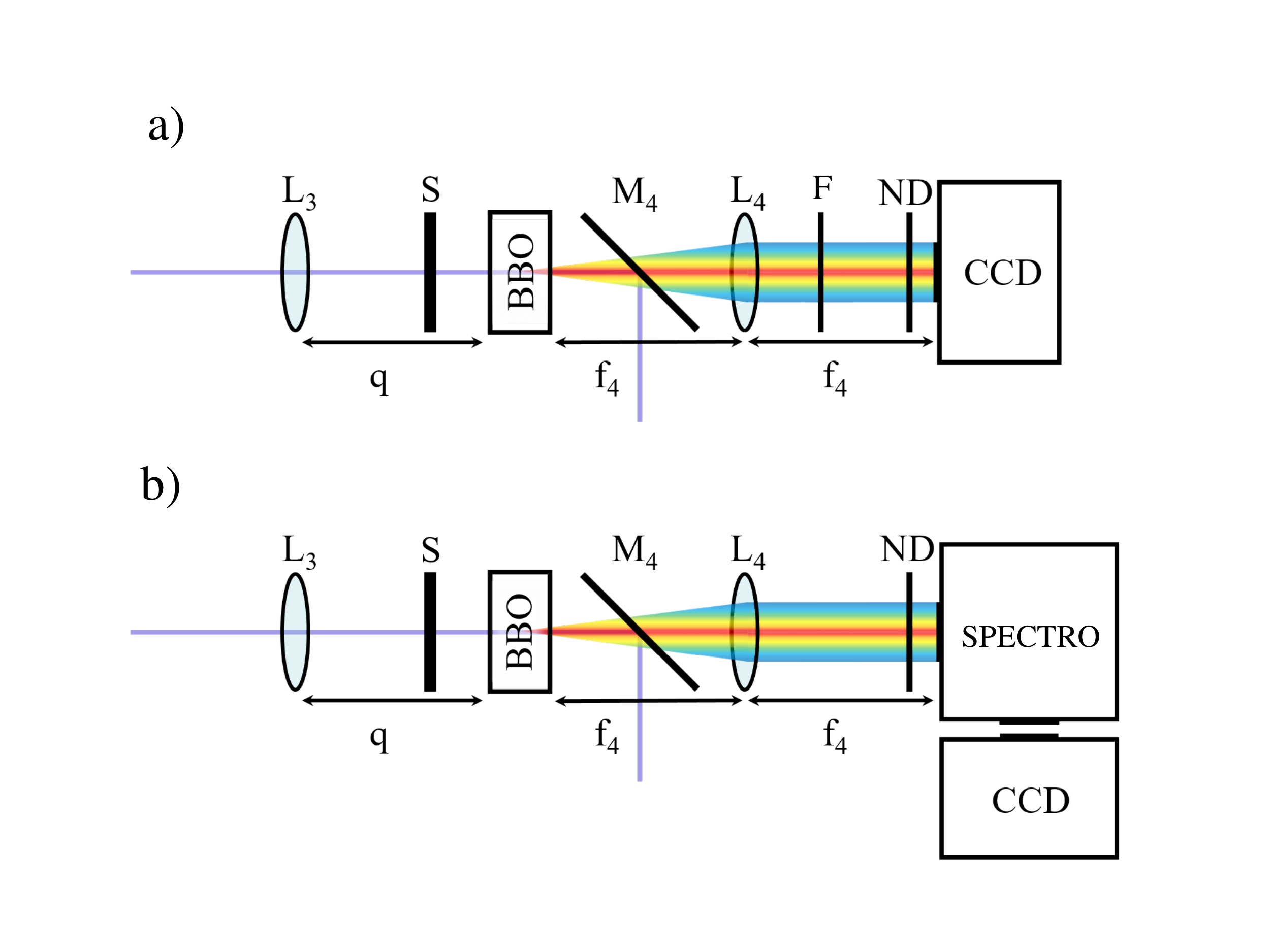}
\caption{Diagnostics schemes: (a) PDC radiation far-field measurements by means of a CCD, and (b) angular-spectral measurements of the radiation by means of an imaging spectrometer coupled to a CCD. The shutter is denoted by S.}
\end{figure}

In the following section we present and analyze the images of the intensity patterns of the PDC radiation generated by a spatially modulated pump and recorded by the CCD as a function of the rotation angle $\beta$ of the BBO crystal in the (xy) plane.
Note that in all measurements the pump beam p$_{1}$ is collinear with the z axis (and thus orthogonal to the input facet of the crystal), in order to maintain the PM surface associated with this pump invariant with $\beta$ (collinear phase-matching). On the other hand the pump beam p$_{2}$ is tilted along the x axis and the PM surfaces associated with this pump vary as function of the crystal orientation in the (xy) plane and the rotation direction. Two different PM situations in the doubly pumped PDC can thus occur; we shall denote them respectively by collinear-non collinear and by collinear-non degenerate phase-matching.

\subsection{Far-field spatial measurements}

A preliminary qualitative observation of the light generated by the BBO crystal pumped in the UV is performed by using a reflex photocamera (Nikon D5100) and taking photographs of the visible pattern of the radiation impinging on a white paper positioned in the far-field of the crystal after a suitable magnification. This allows us to highlight the clear spatial distribution of the generated spectral frequencies in the visible range. In Figure 6 we show an example of a photograph taken in two different pumping configurations, i.e. with a single pump beam p$_{1}$ (Fig. 6a), and with the two interfering pump beams (Fig. 6b) crossing inside the BBO crystal with an angle $\vartheta_{p}\simeq 1.5^{\circ}$; the crystal being oriented for perfect collinear phase-matching for p$_{1}$, and $\beta$ approximately equal to 0$^{\circ}$.  The total pump energy used here is  $E_{p}=55 \pm 3 \mu J$. The pictures have been recorded with an exposure time of 1/4 s and 1/13 s respectively for Fig. 6a and Fig. 6b, the latter case corresponding to one single laser shot and highlighting thus a condition of higher fluence of the emitted radiation in the doubly pumped configuration. We note in Fig. 6b the presence of  coupled modes (lateral bands) and shared modes (central band). Due to the high intensity of the shared and coupled modes, the background PDC fluorescence is not visible in that case.
As expected and discussed in section 2 the angular position of the hot-spots characterizing the \textit{shared} and \textit{coupled} modes clearly depends on the frequency (see Fig.6b).

\begin{figure}[ht]
\centering\includegraphics[width=8cm]{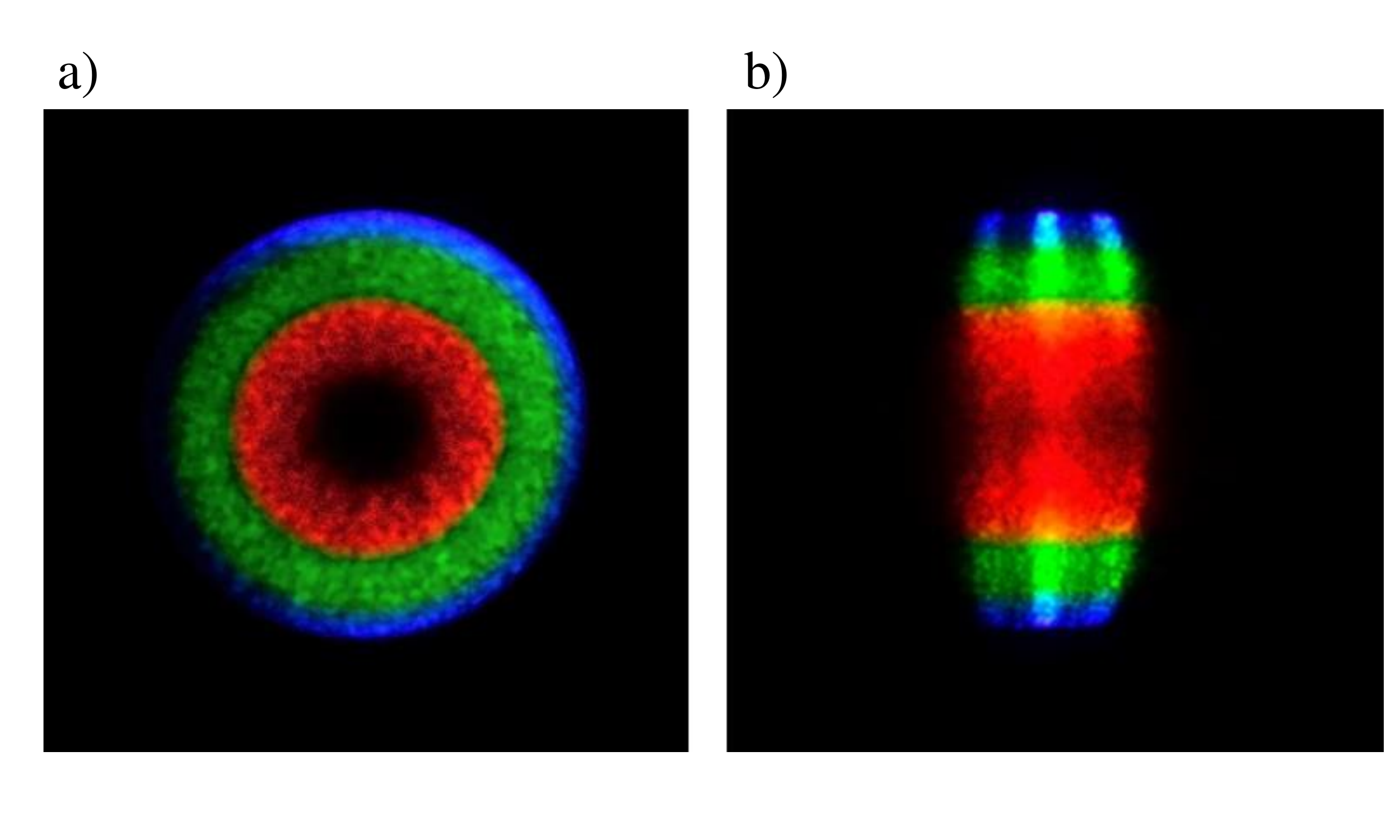}
\caption{Photographs taken by means of a photocamera reflex (nikon) of the far-field intensity distribution of the radiation emitted (a) in the standard single pump PDC process (type I collinear PM)  and (b) in the doubly pumped PDC process for $\vartheta_{p}\simeq 1^{\circ}$ and a rotation angle of the crystal $\beta\simeq 0^{\circ}$. Note that the exposure times of the camera are 1/4 s and 1/13 s respectively in (a) and (b)}
\end{figure}

In order to perform a quantitative analysis, the intensity distribution of the far-field is then recorded with the CCD camera as shown in Fig. 5a. Single shot and multiple shot images are acquired, with an automatic background subtraction performed by the CCD acquisition software. Before the data acquisition the Mach-Zender interferometer has been aligned by optimizing the intensity and the visibility of the three hot-spots bands for $\beta \simeq 0^{\circ}$. In order to identify the condition of \textit{resonance} and the expected intensity enhancement that should characterize a four-mode coupling, the BBO crystal is progressively rotated in the (xy) plane till the reaching of the superposition of the \textit{shared} modes band with one of the \textit{coupled}-modes band. In our experiment for an external angle between the two pumps reasonably equal to 2$^\circ$, the angles $\beta$ corresponding to this configuration turn out to be $\beta \simeq 7^{\circ}$  for the so-called collinear-non collinear PM type case, and $\beta \simeq -8.5^{\circ}$ for the collinear-non degenerate case. The estimated error in the experimental identification of the angle is $1^{\circ}$ given by the reading precision of the rotation stage. These values for $\beta$ corresponding to the so-called "resonance" conditions are in very good accordance with the theoretical values that can be obtained in the plane wave pump approximation,  $7^{\circ}$ and $-8.8^{\circ}$ respectively.

\begin{figure}[ht]
\centering\includegraphics[width=12cm]{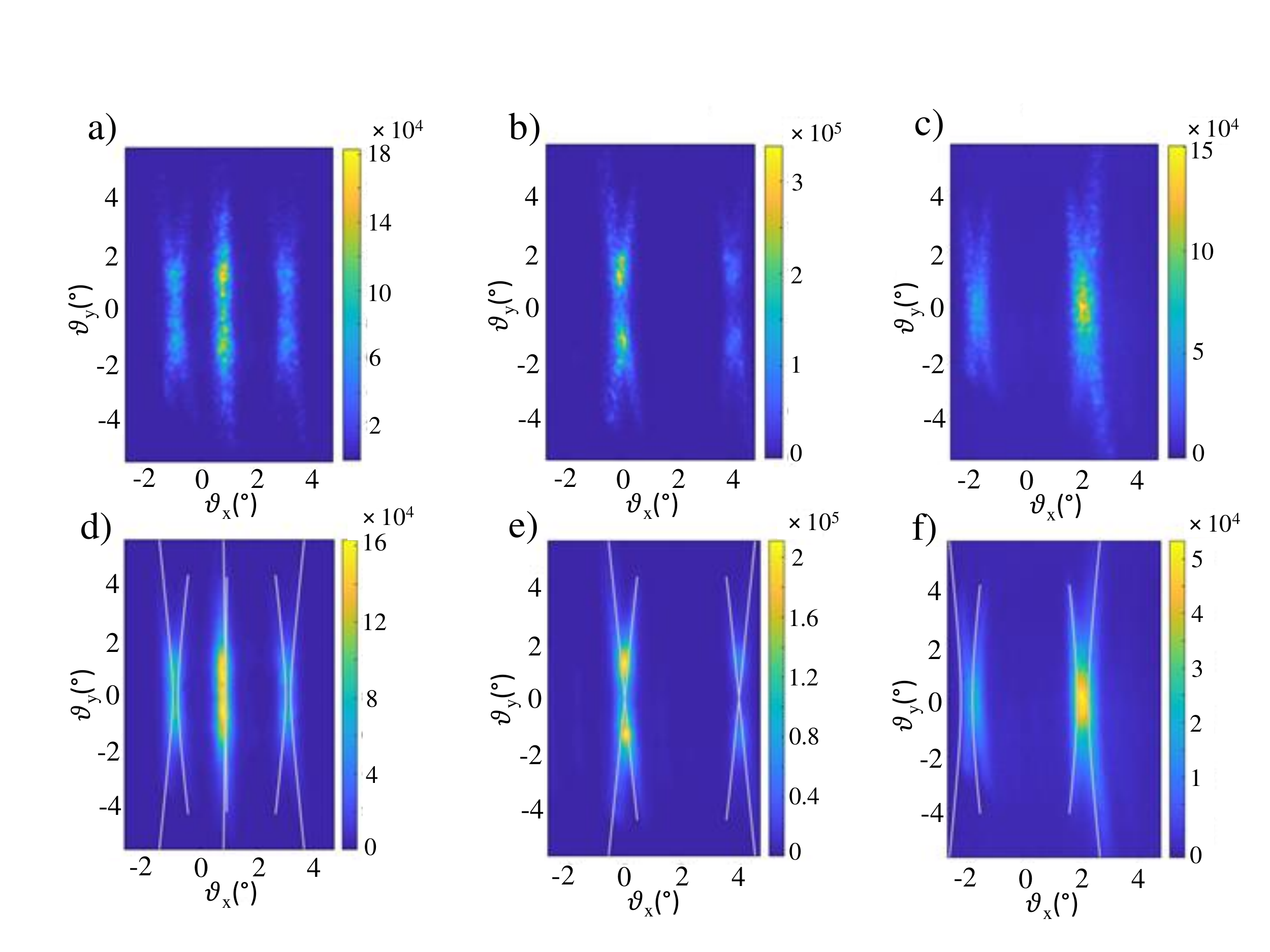}
\caption{Single shot images (top row) and averaged images (bottom row) of the far-field radiation emitted in the doubly pumped PDC process for $\vartheta_{p}\simeq 2^{\circ}$, and for three different crystal rotation angles $\beta = 0^{\circ}$ in (a) and (d), $\beta \simeq 7^{\circ}$ in (b) and (e), and $\beta \simeq -8.5^{\circ}$ in (c) and (f). In (d), (e) and (f) the white curves indicate the theoretical \textit{shared} and \textit{coupled} modes spatial distributions.}
\end{figure} 

In Figure 7 we present single-shot (top row) and averaged (bottom row) images of the far-field radiation emitted in the doubly pumped PDC process, with a total pump energy of $E_{p}\simeq70 \pm 3 \mu J$. The bottom-row images are obtained by averaging 100 single shot acquisitions.
Given the fact that the separation distance between the two external hot-spots bands associated with the \textit{coupled} modes corresponds to the double of the separation angle $\theta_{p}$ between the two pumps, these results confirm an angle of $\vartheta_{p} \simeq 2^{\circ}$. Considering this value and the pumping configuration (one pump collinear with the z axis and one tilted pump along x in the (xz) plane), we have superposed the theoretical curves of the \textit{shared} and \textit{coupled} mode spatial distributions in the far field onto the averaged intensity images (Fig.7d-f), obtaining an excellent agreement with the experimental data. It is also worth noting in the background the presence of very weak fluorescence radiation cones generated by each single pump.

\subsection{Angular-spectral measurements}

In order to study the spectral content of the generated hot-spots and to quantitatively characterize their spatial distribution as a function of the wavelength, we have recorded the spectrally resolved far-field images of the PDC radiation by using the imaging spectrometer and the CCD camera, as shown in Fig. 5b. We have chosen to select the section $\theta_{y}=0^{\circ}$ with the slit, and thus analyze the spectral distribution along $\theta_{x}$ (i.e. in the horizontal plane, containing the two pump beams). These spectra corresponding to the doubly pumped PDC configuration with  $\vartheta_{p}=2^{\circ}$, have been recorded again for three different values of the crystal rotation angle $\beta$.

\begin{figure}[ht]
\centering\includegraphics[width=12cm]{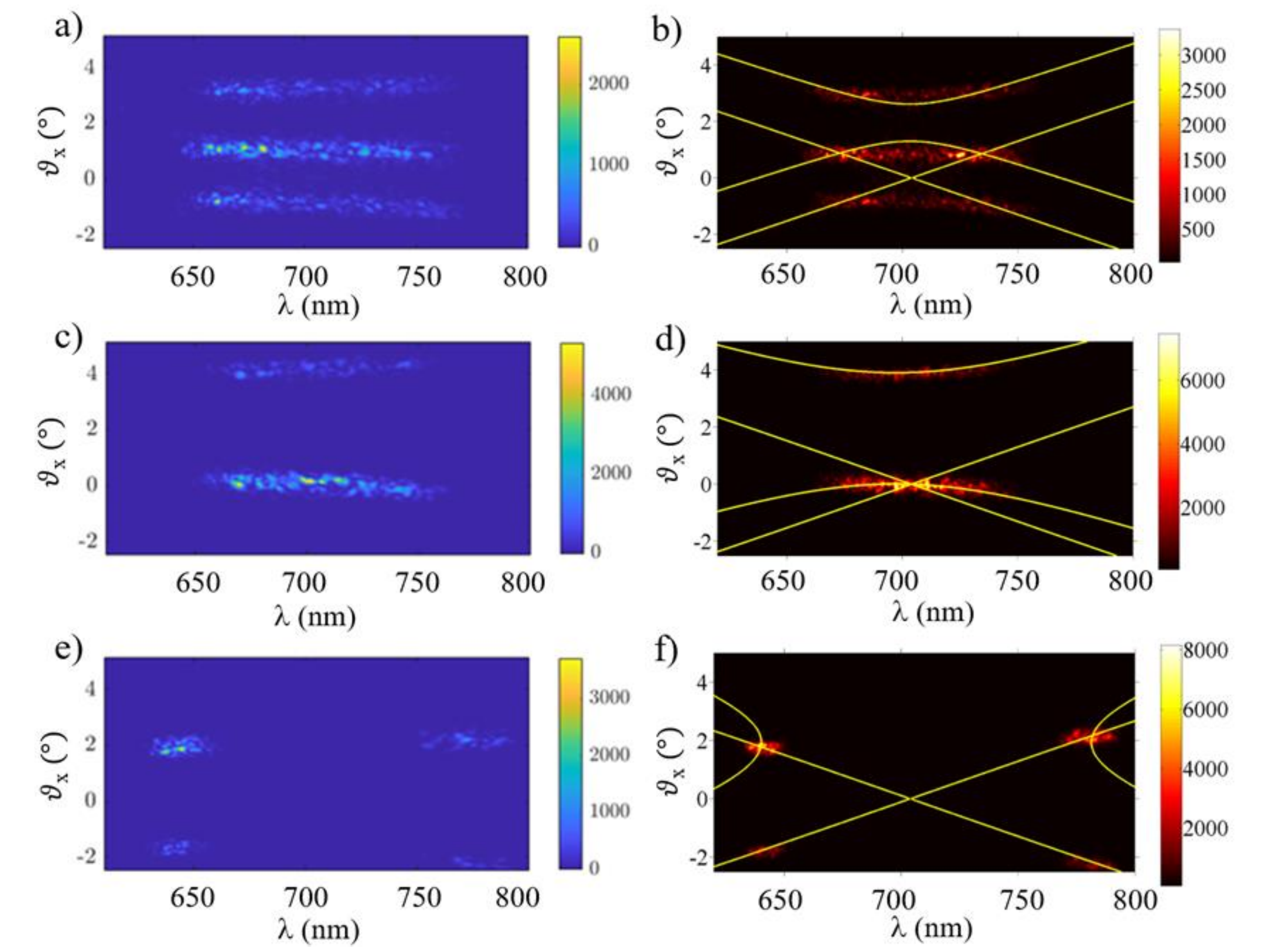}
\caption{Example of spectral distributions of the PDC radiation far-field selected by the interferometer slit in the horizontal plane ($\vartheta_{y}=0^{\circ}$). First column: recorded single shot far-field spectra; second column: results of the numerical simulations of single shot far-field spectra, with a crystal gain coefficient per length $g$ such that $gl_{c}$=6 with a crystal length $l_{c}$=4mm.  The images correspond to a crystal rotation angle of $\beta = 0^{\circ}$ in (a,b), $\beta \simeq 7^{\circ}$ in (c,d), and $\beta \simeq -8.5^{\circ}$ in (e,f). The superimposed yellow curves are the calculated PM curves associated with the two pumps in the three cases respectively.}
\end{figure}

The images reported in the first column of Figure 8 are single shot images corresponding to a total pump energy of about $E_{p}\simeq 62 \pm 4 \mu J$, while those in the second column correspond to the results of numerical simulations, in the three cases, namely $\beta=0^{\circ}$ (a,b), $\beta=7^{\circ}$ (c,d) and $\beta=-8.8^{\circ}$ (e,f).
Note that experimentally the parametric fluorescence of the PDC is not visible (unless we truncate the images counts at a low level) due to the stronger intensity  of the \textit{shared} and \textit{coupled} modes with respect to the background. In contrast to the experiment realized with the hexagonally poled crystal \cite{Jedrkiewicz2018}, here the \textit{shared} and \textit{coupled} modes do not appear as very localized hot-spots in the ($\lambda$,  $\vartheta_{x}$) plane, but as extended stripes. This could be due both to the different PM conditions, especially for what concerns Fig. 8a and b, but also to the much shorter length of the crystal in the BBO case, which results in shared modes distributed over a wider range of frequencies. The numerical simulation results are in excellent agreement with our experimental results, and highlight the fact that the high intensity stripes are not parallel one to another, and thus that the radiation spatial distribution (i.e. the emission position $\vartheta_{x}$) has a frequency dependence also in the horizontal plane ($\vartheta_{y}=0^{\circ}$).

\begin{figure}[ht]
\centering\includegraphics[width=12cm]{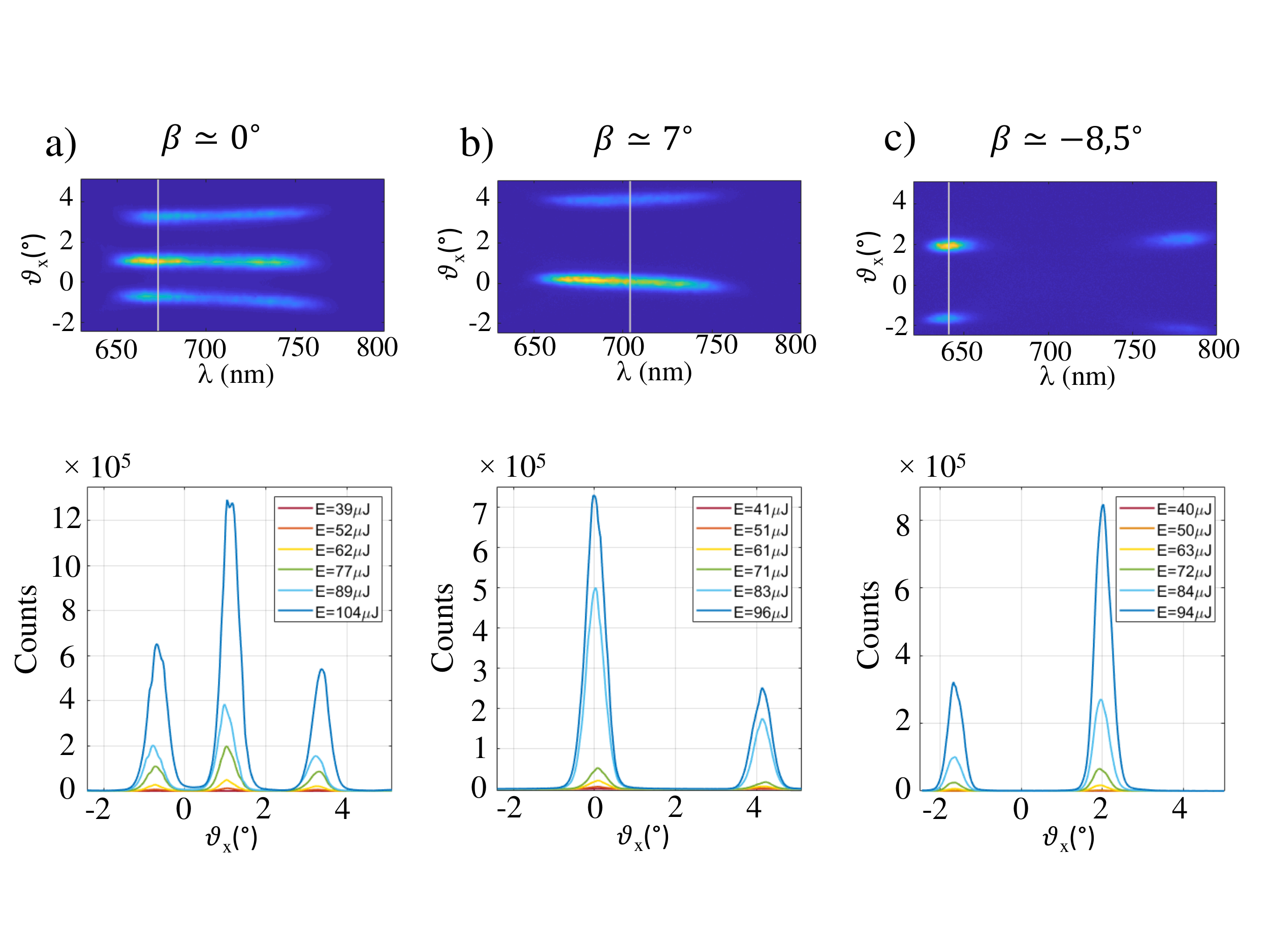}
\caption{Top row: Example of averaged spectra of the radiation emitted in the doubly pumped PDC process (pump energy 62 $\pm$ 4 $\mu J$). Bottom row, sections taken along $\vartheta_{x}$ in different averaged spectral images recorded for different values of the total pump energy; the profiles are taken for the wavelengths corresponding to the \textit{shared} modes, as indicated  by the gray lines drawn over the images.}
\end{figure}

In figure 9 we report for illustration, experimental averaged spectra performed over 150 single-shot recorded images for the same energy value as before. Transverse sections along $\theta_{x}$ are also analyzed in various averaged spectral images recorded for different values of the total pump energy (equally distributed among the two pumps). They are plotted in correspondence to the theoretical wavelengths of the \textit{shared} modes (those that one would have in the case of an infinite length crystal), as indicated for instance by the gray lines superposed on the images in the top row of Fig. 9. In particular if $\beta=0^{\circ}$ (Fig. 9a) one finds $\lambda_{s}=673$ nm for the \textit{shared} signal and $\lambda_{s}=738$ nm for the \textit{shared} idler. In the case of $\beta=7^{\circ}$ (Fig. 9b), the emission of the \textit{shared} modes occurs at degeneracy with $\lambda_{s,i}=704$ nm, while for $\beta=-8.8^{\circ}$ (Fig.9c), one has $\lambda_{s}=641$ nm for the \textit{shared} signal and $\lambda_{s}=781$ nm for the \textit{shared} idler. In particular given the higher CCD detection efficiency for higher frequencies, we have chosen to concentrate our attention on the wavelengths associated with the signal fields even if the \textit{shared} and \textit{idler} modes belonging to the considered section are not coupled among them and do not belong to the same triplet of modes. The image sections presented in Figure 9 are obtained after a count integration centered around $\lambda_{s}$ over a 5 nm wavelength interval (corresponding to 22 pixels), in the three crystal orientation cases. As expected, we see in Fig. 9a that the peak intensity of the \textit{shared} modes is about the double of that of the \textit{idler} modes.
In the cases of Fig. 9b and c (four mode coupling scenario) in analogy to what described in \cite{Jedrkiewicz2018,Gatti2018}, the mean ratio between the maximum counts of the \textit{shared} and \textit{coupled} modes has been evaluated to be $3.11\pm0.25$ for the configuration with $\beta=7^{\circ}$, and $2.64\pm0.05$ for $\beta=-8.5^{\circ}$, close to and in very good accordance respectively with the theoretical prediction in the plane wave pump approximation given by $\phi^{2}=2.618$, where $\phi$ is the Golden Ratio \cite{Gatti2018}.

\subsection{Gain enhancement}

The parametric gain enhancement in the different configurations of the experiment, i.e. for different crystal orientations, can be studied by looking at the different behaviour of the radiation intensity as a function of the square root of the total pump energy $\sqrt{E_{p}}$. This can be analyzed in a logarithmic scale, exploiting the fact that in the stimulated regime the intensity undergoes an exponential grow as function of $\sqrt{E_{p}}$ \cite{Jedrkiewicz2018}.
To this end we have looked at the evolution of the counts per pixel selected in suitable regions of the recorded spectra in different energy pumping regimes.  This study has been performed with averaged images as those reported in the top row of Figure 9. We have integrated the counts number of the \textit{shared} modes over a rectangular area of 1000 pixels around the corresponding central wavelengths $\lambda_{s}$, in particular $\lambda_{s}=673$ nm in the case of $\beta=0^{\circ}$, $\lambda_{s}=704$ nm for $\beta=7^{\circ}$, and $\lambda_{s}=641$ nm for $\beta=-8.5^{\circ}$. For a comparison with the standard two mode coupling, we have managed to select in the spectral images an area of 900 pixels along the tails of the background PDC fluorescence from a single pump, detectable in the configuration of non-rotated crystal.

\begin{figure}[ht]
\centering\includegraphics[width=12cm]{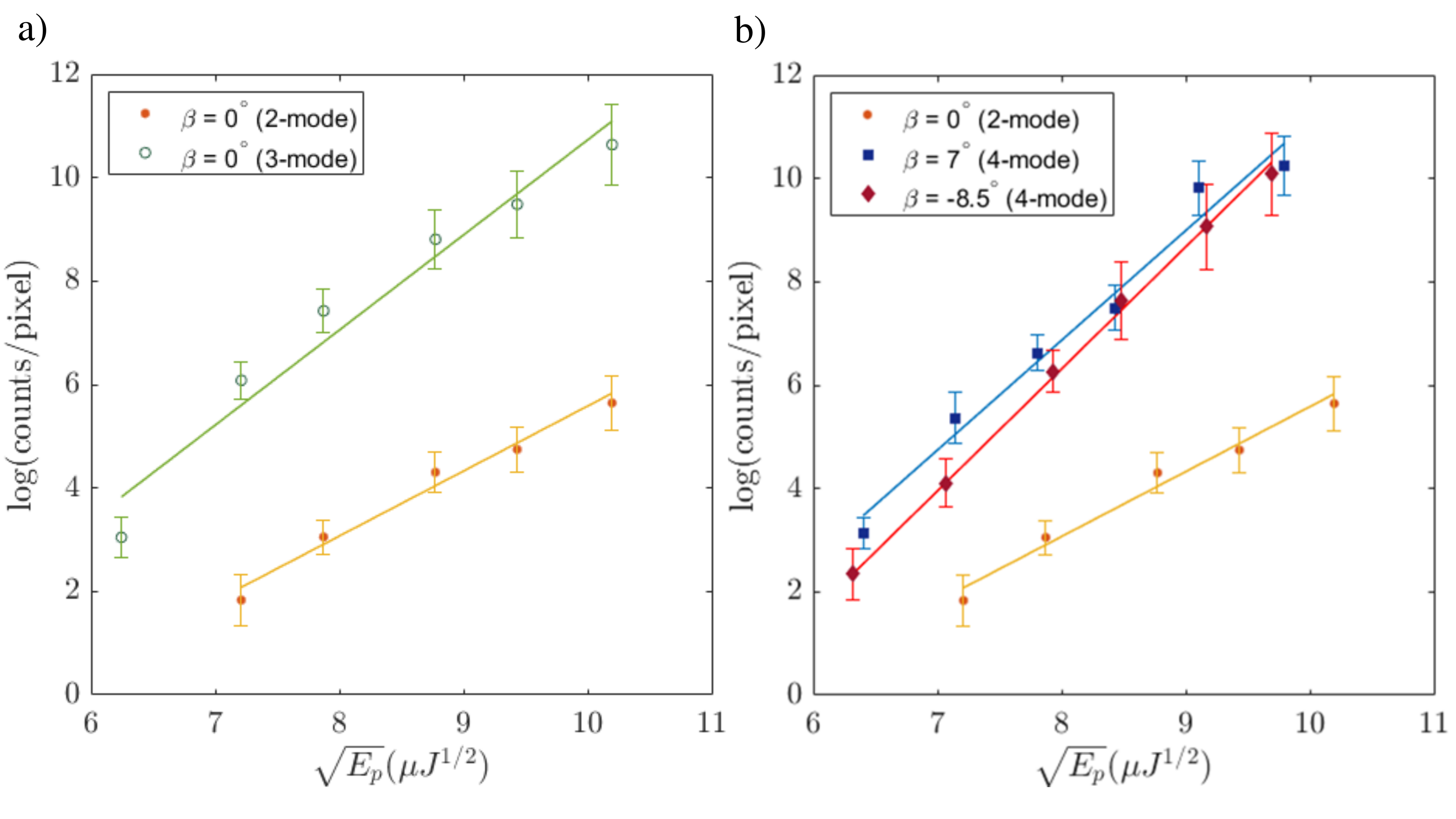}
\caption{Gain enhancement of the coherently coupled processes in the doubly pumped PDC process. \textit{Shared}-modes photon numbers plotted in logarithmic scale versus the square-root of the pump energy. The markers show experimental
data in the different coupling regimes, while the lines are the corresponding linear fits. Three-mode and four-mode coupling regimes are compared with the two-mode standard PDC fluorescence respectively in (a) and (b).}
\end{figure}

In Figs. 10a  and 10b, the points represent the experimental data extracted from the spectra ($\lambda$, $\theta_{x}$) in different regimes, averaged over 150 images and recorded for different energy pump values. The error bars evaluated by means of the error propagation technique reflect both the standard deviation associated with the mean counts per pixel in the selected area, and the pump energy fluctuations. 
From the ratio between the angular coefficient of the line fit corresponding to the shared modes where a three-mode coupling is realized and that of the background fluorescence corresponding to a standard two-mode coupling,
we find a gain enhancement of $1.46\pm0.29$ for $\beta=0^\circ$ perfectly in accordance with the value $\sqrt{2}$ predicted by the plane wave pump theory showing that $I_{3modes}=(I_{2modes})^{\sqrt{2}}$.
We also find a gain enhancement of $1.68\pm0.29$ and  $1.87\pm0.26$ for the cases respectively associated with $\beta=7^{\circ}$ and $\beta=-8.5^{\circ}$, both values being compatible with the theoretical expectation $\phi=1.618$, having $I_{4modes}=(I_{2modes})^{\phi}$.

\section{Conclusion}
In this work we have studied the parametric down-conversion process in a nonlinear BBO crystal, stimulated by the interference pattern of two pump beams, obtained from a Mach-Zender type interferometer. We have shown the possibility to reproduce with a bulk crystal a similar phenomenology to that observed previously in an experiment of PDC with a 2D nonlinear photonic crystal. The spatial far-field distributions of the \textit{shared} and \textit{idler} modes of the generated radiation, featured by hot spots, in accordance to what predicted by the theory, have been clearly observed in the experiment. A transition from a three-mode coupling to a four-mode coupling has been reached by exploiting the crystal anisotropy and by rotating the crystal input face in such a way to introduce a refractive index change of one of the two pumps, and thus a wave number variation inside the crystal. In a parallel work \cite{Gatti2020} we give an interpretation of this transition in terms of a superposition of the career Poynting vector, which identifies the direction of propagation of the energy flux, with either one pump mode or the other.
The spectrally resolved far field patterns recorded have confirmed that the resonance condition can occur for a large bandwidth around the degenerate wavelength, and that a gain enhancement is present in these special hot modes regions. This experiment has highlighted that the possibility to tune the angular-spectral distributions of the \textit{shared} and \textit{idler} modes of the PDC pattern, by simply modifying the geometry of the pumping conditions and the crystal orientation, make simple bulk nonlinear sources interesting and reliable for the generation of particular ad hoc multimode states, featured as predicted by the theory by coupling and entanglement that might be used for different applications.

\section*{Disclosures}

Disclosures. The authors declare no conflicts of interest.




\end{document}